\documentclass[aps,prl,reprint,twocolumn,superscriptaddress,floatfix,nofootinbib,longbibliography]{revtex4-1}
\usepackage{graphicx,amsmath,amsfonts,amssymb,amsthm,xr}
\usepackage{epsfig,amsmath,amssymb,color,dsfont,upgreek,physics}
\usepackage{mathrsfs}
\usepackage{mathtools}
\usepackage{bbold}
\usepackage{comment}
\usepackage{float}

\usepackage[bookmarks=true,colorlinks,linkcolor=OrangeRed,urlcolor=NavyBlue,citecolor=RoyalBlue]{hyperref}
\usepackage[dvipsnames]{xcolor}
\usepackage{orcidlink}

\definecolor{mygold}{rgb}{0.93,0.69,0.13}

\definecolor{mypurple}{rgb}{0.49,0.18,0.56}

\definecolor{mygreen}{rgb}{0,0.5,0}

\definecolor{mygreen}{rgb}{0,0.5,0}
\definecolor{myred}{rgb}{0.7,0,0}
\definecolor{myblue}{rgb}{0,0,1}

\begin{document}
\title{Large-Scale $2+1$D $\mathrm{U}(1)$ Gauge Theory with Dynamical Matter in a Cold-Atom Quantum Simulator}
\author{Jesse Osborne${}^{\orcidlink{0000-0003-0415-0690}}$}
\affiliation{School of Mathematics and Physics, The University of Queensland, St. Lucia, QLD 4072, Australia}
\author{Ian P.~McCulloch${}^{\orcidlink{0000-0002-8983-6327}}$}
\affiliation{School of Mathematics and Physics, The University of Queensland, St. Lucia, QLD 4072, Australia}
\author{Bing Yang${}^{\orcidlink{0000-0002-8379-9289}}$}
\email{yangbing@sustech.edu.cn}
\affiliation{Department of Physics, Southern University of Science and Technology, Shenzhen 518055, China}
\author{Philipp Hauke${}^{\orcidlink{0000-0002-0414-1754}}$}
\email{philipp.hauke@unitn.it}
\affiliation{INO-CNR BEC Center and Department of Physics, University of Trento, Via Sommarive 14, I-38123 Trento, Italy}
\affiliation{INFN-TIFPA, Trento Institute for Fundamental Physics and Applications, Trento, Italy}
\author{Jad C.~Halimeh${}^{\orcidlink{0000-0002-0659-7990}}$}
\email{jad.halimeh@physik.lmu.de}
\affiliation{Department of Physics and Arnold Sommerfeld Center for Theoretical Physics (ASC), Ludwig-Maximilians-Universit\"at M\"unchen, Theresienstra\ss e 37, D-80333 M\"unchen, Germany}
\affiliation{Munich Center for Quantum Science and Technology (MCQST), Schellingstra\ss e 4, D-80799 M\"unchen, Germany}

\begin{abstract}
A major driver of quantum-simulator technology is the prospect of probing high-energy phenomena in synthetic quantum matter setups at a high level of control and tunability. Here, we propose an experimentally feasible realization of a large-scale $2+1$D $\mathrm{U}(1)$ gauge theory with dynamical matter and gauge fields in a cold-atom quantum simulator with spinless bosons. We present the full mapping of the corresponding Gauss's law onto the bosonic computational basis. We then show that the target gauge theory can be faithfully realized and stabilized by an emergent gauge protection term in a two-dimensional single-species Bose--Hubbard optical Lieb superlattice with two spatial periods along either direction, thereby requiring only moderate experimental resources already available in current cold-atom setups. Using infinite matrix product states, we calculate numerical benchmarks for adiabatic sweeps and global quench dynamics that further confirm the fidelity of the mapping. Our work brings quantum simulators of gauge theories a significant step forward in terms of investigating particle physics in higher spatial dimensions, and is readily implementable in existing cold-atom platforms.
\end{abstract}

\date{\today} 
\maketitle 
\textbf{\textit{Introduction.---}}Gauge theories serve as a fundamental framework of modern physics through which various outstanding phenomena such as quark confinement, topological spin liquid phases, high-$T_\text{c}$ superconductivity, and the fractional quantum Hall effect can be formulated \cite{Cheng_book,Balents_NatureReview,Savary2016}. This makes gauge theories relevant across various fields ranging from condensed matter to high-energy physics. Gauge theories are characterized by their principal property of \textit{gauge invariance}, which is a local symmetry that enforces an intrinsic relation between local sets of degrees of freedom, and gives rise to gauge fields, the mediators of interactions between elementary particles \cite{Zee_book}. The Standard Model of particle physics includes prominent gauge theories such as quantum electrodynamics, with its Abelian $\mathrm{U}(1)$ gauge symmetry, and quantum chromodynamics, with its non-Abelian $\mathrm{SU}(3)$ gauge symmetry \cite{Weinberg_book,Gattringer_book}.

With the great progress achieved in the control and precision of modern quantum simulators \cite{Greiner2002,Bloch2008,Bakr2009,Trotzky2011,Hauke2012,Georgescu_review}, recent years have witnessed a tremendous interest in realizing gauge theories in various synthetic quantum matter platforms such as, e.g., superconducting qubits, Rydberg setups, and optical lattices  \cite{Pasquans_review,Dalmonte_review,Zohar_review,aidelsburger2021cold,Zohar_NewReview,Bauer_review}. Such setups can serve as an experimental probe that can ask questions complementary to dedicated classical computations and high-energy colliders, with the exciting potential to calculate time evolution from first principles \cite{Bauer_review}. Until now, almost all quantum-simulation experiments of gauge theories have taken place in one spatial dimension or have been restricted to a small number of plaquettes \cite{Bernien2017,Kokail2019,Martinez2016,Muschik2017,Klco2018,Schweizer2019,Goerg2019,Mil2020,Klco2020,Yang2020,Zhou2022,Nguyen2021,Wang2021,Mildenberger2022,Wang2022}. Although these experiments are significant milestones in their own right, going to higher spatial dimensions is essential for probing salient high-energy phenomena in nature. Developing scalable implementations for higher dimensions has thus been identified as a major challenge in the field \cite{Zohar_NewReview}.

Here, we propose an experimentally feasible quantum simulator of a large-scale $2+1$D $\mathrm{U}(1)$ gauge theory with dynamical matter and gauge fields. This is achieved by mapping the model onto spinless bosonic degrees of freedom on an optical superlattice; see Fig.~\ref{fig:setup}(a). We provide perturbation theory derivations outlining the stability of the gauge symmetry in this mapping due to an emergent \textit{linear gauge protection term} \cite{Halimeh2020e,Lang2022stark}, and perform time-evolution numerical benchmarks using infinite matrix product state (iMPS) techniques that demonstrate robust fidelity of the mapping. Our work complements previous proposals for implementing gauge theories in higher dimensions, which mostly concentrated on pure gauge theories (see, e.g., \cite{Buechler2005,Zohar2011,Tagliacozzo2013,Dutta2017,Ott2020scalable,Fontana2022}), while here we include dynamical matter (see also \cite{Zohar2013,Paulson2020simulating,Homeier2022quantum}), but without the use of a plaquette term.
Recent theory investigations have
demonstrated that even in the absence of plaquette terms, $2+1$D gauge theories with dynamical matter can nevertheless display extremely rich physics, such as symmetry protected topological states and spin liquid phases \cite{Cardarelli2017,Ott2020Noncancellation,Cuadra2020,Hashizume2022}. Importantly, our proposal is feasible to implement with existing technology in current cold-atom experiments.

\textbf{\textit{Model and mapping.---}}We consider a quantum link formulation of the $2+1$D Abelian Higgs model, where, in keeping with experimental feasibility, the infinite-dimensional gauge and electric fields are represented by spin-$1/2$ operators \cite{Chandrasekharan1997,Wiese_review}. The resulting $2+1$D $\mathrm{U}(1)$ quantum link model (QLM) is described by the Hamiltonian
\begin{align}\label{eq:QLM}
    \hat{H}_\mathrm{QLM}{=}{-}\kappa\sum_{\mathbf{r},\nu}\Big(\hat{\phi}_\mathbf{r}^\dagger\hat{s}^-_{\mathbf{r},\mathbf{e}_\nu}\hat{\phi}_{\mathbf{r}+\mathbf{e}_\nu}^\dagger{+}\text{H.c.}\Big){+}m\sum_\mathbf{r}\hat{\phi}^\dagger_\mathbf{r}\hat{\phi}_\mathbf{r},
\end{align}
where $\mathbf{r}{=}\big(r_x,r_y\big)^\intercal$ is the vector specifying the position of a lattice site, and $\mathbf{e}_\nu$ is a unit vector along the direction $\nu{\in}\{x,y\}$, with the lattice spacing set to unity throughout this work. The hard-core bosonic ladder operators $\hat{\phi}_\mathbf{r},\hat{\phi}_\mathbf{r}^\dagger$ act on the matter field at site $\mathbf{r}$ with mass $m$, with $\kappa$ the coupling strength, while the spin-$1/2$ operators $\hat{s}^\pm_{\mathbf{r},\mathbf{e}_\nu}$ and $\hat{s}^z_{\mathbf{r},\mathbf{e}_\nu}$ represent the gauge and electric fields, respectively, at the link between sites $\mathbf{r}$ and $\mathbf{r}{+}\mathbf{e}_\nu$. Furthermore, we have adopted a particle-hole transformation \cite{Hauke2013} that allows for a more intuitive connection to the bosonic mapping employed in this work---see Fig.~\ref{fig:setup} and Supplemental Material (SM) for details \cite{SM}.

The generator of the $\mathrm{U}(1)$ gauge symmetry of Hamiltonian~\eqref{eq:QLM} is
\begin{align}\label{eq:Gr}
    \hat{G}_\mathbf{r}=(-1)^{r_x+r_y}\bigg[\hat{\phi}^\dagger_\mathbf{r}\hat{\phi}_\mathbf{r}{+}\sum_{\nu}\Big(\hat{s}^z_{\mathbf{r},\mathbf{e}_\nu}{+}\hat{s}^z_{\mathbf{r}-\mathbf{e}_\nu,\mathbf{e}_\nu}\Big)\bigg],
\end{align}
which is equivalent to a discretized version of Gauss's law, where $\big[\hat{G}_\mathbf{r},\hat{G}_{\mathbf{r}'}\big]=0$. The gauge invariance of Hamiltonian~\eqref{eq:QLM} is encoded in the commutation relations $\big[\hat{H}_\text{QLM},\hat{G}_\mathbf{r}\big]=0,\,\forall\mathbf{r}$. The \textit{physical sector} of Gauss's law is the set of gauge-invariant states $\{\ket{\psi}\}$ satisfying $\hat{G}_\mathbf{r}\ket{\psi}=0,\,\forall\mathbf{r}$. This restricts the allowed configurations of matter on a given site and the electric fields on its four neighboring links to those depicted in Fig.~\ref{fig:setup}(b), where we also show the corresponding configurations in the bosonic model onto which we map the QLM in order to quantum-simulate it. The crux of this mapping lies in first restricting the local Hilbert space of each ``matter'' site in the BHM superlattice to two states: $\ket{0}_\mathbf{r}$ and $\ket{1}_\mathbf{r}$, which correspond to an empty or occupied matter field on that site in the QLM. Furthermore, we restrict the local Hilbert space of each ``gauge'' site to the states $\ket{0}_{\mathbf{r},\mathbf{e}_\nu}$ and $\ket{2}_{\mathbf{r},\mathbf{e}_\nu}$, where an empty gauge site corresponds to the local electric field pointing left (down) on the corresponding horizontal (vertical) link in the QLM, while a doublon on the gauge site of the bosonic model denotes a local electric field pointing right (up).

\begin{figure}[t!]
	\centering
    \includegraphics[width=\linewidth]{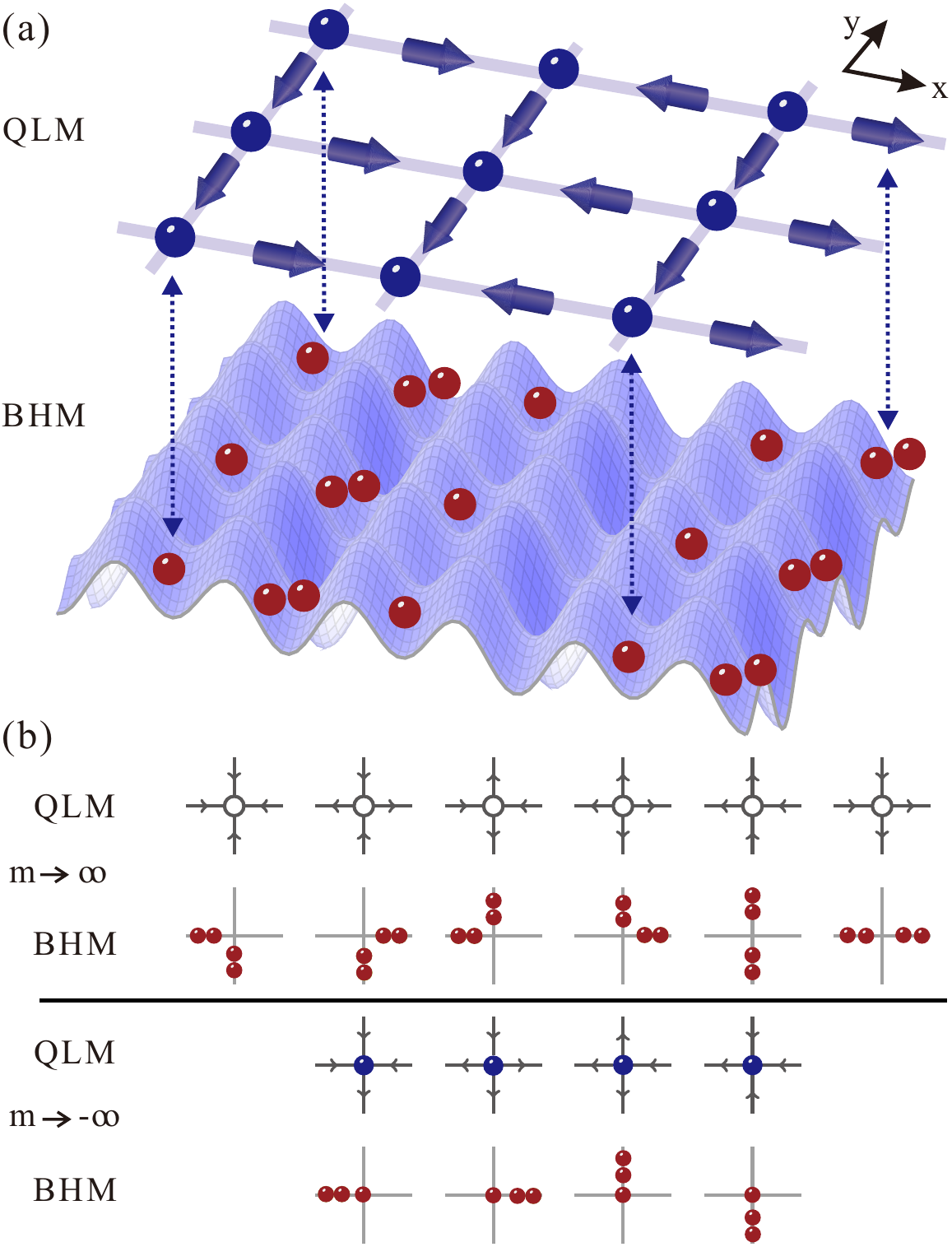}
	\caption{(Color online). (a) Mapping of the $2+1$ $\mathrm{U}(1)$ quantum link model onto the two-dimensional Bose--Hubbard model on a Lieb superlattice. A matter site (shallow) can host one or zero bosons, while a gauge site (deep) can host two or zero bosons. Forbidden sites (deepest) are never occupied. (b) The different matter and electric-field configurations permitted by Gauss's law in the computational basis, shown for both the QLM and its BHM mapping.}
	\label{fig:setup}
\end{figure}

Such local constraints can be achieved by employing the $2$D tilted Bose--Hubbard model (BHM) on a Lieb superlattice, given by the Hamiltonian
\begin{align}\nonumber
\hat{H}_\text{BHM}=\sum_\mathbf{j}&\bigg[J\sum_{\nu=x,y}\Big(\hat{b}_\mathbf{j}^\dagger\hat{b}_\mathbf{j+\mathbf{e}_\nu}+\text{H.c.}\Big)\\\label{eq:BHM}
&+\frac{U_\mathbf{j}}{2}\hat{n}_\mathbf{j}\big(\hat{n}_\mathbf{j}-1\big)+\big(\vec{\gamma}^\intercal\mathbf{j}-\delta_\mathbf{j}-\eta_\mathbf{j}\big)\hat{n}_\mathbf{j}\bigg],
\end{align}
where $\mathbf{j}{=}(j_x,j_y)^\intercal$ is a vector denoting the position of a site on the square superlattice, $\hat{b}_\mathbf{j}$ and $\hat{b}_\mathbf{j}^\dagger$ are bosonic ladder operators satisfying the canonical commutation relations $\big[\hat{b}_\mathbf{j},\hat{b}_\mathbf{l}\big]{=}0$ and $\big[\hat{b}_\mathbf{j},\hat{b}_\mathbf{l}^\dagger\big]{=}\delta_{\mathbf{j},\mathbf{l}}$ with $\hat{n}_\mathbf{j}{=}\hat{b}_\mathbf{j}^\dagger\hat{b}_\mathbf{j}$ the singlon number operator, $\vec{\gamma}{=}(\gamma_x,\gamma_y)^\intercal$ encodes the tilt due to a magnetic gradient potential in each direction, and the chemical potential $\delta_\mathbf{j}$ ($\eta_\mathbf{j}$) equals $\delta$ ($\eta$) only on link (forbidden) sites and zero elsewhere; see Fig.~\ref{fig:setup}(a). The on-site interaction strength $U_\mathbf{j}$ is $U$ on gauge sites and $\alpha U$ on matter sites, where $\alpha{\approx}1.3$ (see below for experimental details).

The QLM Hamiltonian~\eqref{eq:QLM} can be derived from the BHM Hamiltonian~\eqref{eq:BHM} as a leading effective theory in second-order perturbation theory in the regime $\eta\gg U\sim2\delta\gg J,m$, where the parameters of both models can be related as $\kappa\approx4\sqrt{2}J^2/U$ and $m=\delta-U/2$ \cite{SM}. 
It is important to note here that for a fixed value of $m$ in the BHM a slightly renormalized effective mass $\tilde{m}$ arises in the $\mathrm{U}(1)$ QLM due to ``undesired'' gauge-invariant second-order processes \cite{SM}. Nevertheless, this renormalization does not break gauge symmetry, and the simulated model is still $\mathrm{U}(1)$ gauge-invariant. Indeed, as we detail in the SM \cite{SM}, our BHM quantum simulator hosts a $\mathrm{U}(1)$ gauge symmetry that is stabilized through the concept of \textit{linear Stark gauge protection} \cite{Halimeh2020e,Lang2022stark}.

\textbf{\textit{Experimental setup.---}}The BHM Hamiltonian~\eqref{eq:BHM} can be faithfully realized using cold atoms. Indeed, the $2+1$D $\mathrm{U}(1)$ QLM can be mapped and implemented using the technology of the $1+1$D state-of-the-art Bose--Hubbard quantum simulator \cite{Yang2020,Zhou2022}.
Starting from degenerated quantum gases, ultracold bosons trapped in the $2$D optical lattices are described by the BHM. Here, the tunnelling strength $J$ and the on-site interaction $U$ are controlled primarily by tuning the depth of the optical lattices. When one slowly increases the ratio $U/J$ to approach the atomic limit at $U/J \gg 1$, the system undergoes a superfluid-to-Mott insulator phase transition. Other than the general Bose--Hubbard settings, our lattice gauge theory model poses strong constraints on the quantum states of the atoms. To truncate the system into the gauge-allowed subspace, we must initialize the atom occupation to fulfill Gauss's law of Eq.~\eqref{eq:Gr} and meanwhile control the optical lattices to prevent gauge-violating processes.

Here, we propose a feasible way to realize the model using ultracold bosons in a special $2$D superlattice \cite{Dai2017}. The potential of the bichromatic superlattice can be written as $V(r)= V_s(r)\cos^2(4\pi r/\lambda) - V_l(r)\cos^2(2\pi r/\lambda - \pi/4)\ \text{with} \ r=\{x,y\}$, where $\lambda$ is the wavelength of the `short' lattice, and $V_{s,l}$ are the lattice depths of the `short' and `long' lattices, respectively.
The overlapping of the intensity minima of the lattices enables the largest imbalance between the neighboring lattice sites.
The initial state shown in Fig.~\ref{fig:ramp}(a) at $m \rightarrow +\infty$ is a special type of Mott insulating state.
It can be obtained by first cooling the quantum gases in superlattices \cite{Yang:2020Science} and then following up with a selective atom-removing operation.
When the average filling factor is set to $\bar{n} = 0.625$, and the lattice depth is $V_l(x)=U/2, \ V_l(y) \gg U$ at the end stage of the Mott transition, the atoms reside only on even columns and the mean filling factor is $\ket{...,2,0.5,2,0.5,2,0.5,...}$.
Next, one could remove the atoms residing on the matter sites and achieve a clear initial state \cite{Yang:2017pra}.

Furthermore, we suggest tuning the ratio $\alpha$ of the on-site interactions using the difference between the Wannier states of two Bloch bands.
When $V_l$ is set close to the band gap ($s$ to $p$-band) of the short lattices, the atoms can live on the $s$-band of the matter sites but only on the $p$-band of the gauge sites.
The ratio is $\alpha\approx1.3$ in the experimentally relevant regime. In this sense, the energy shifts $\delta_{\mathbf{j}}$ and $\eta_\mathbf{j}$ denote the energy difference between the matter-site $s$-band and the gauge-site $p$-band. Such a superlattice structure forms a very deep potential on the forbidden sites, shifting the energy levels away from resonance.
To further prevent atoms from tunnelling to the forbidden sites, a larger energy penalty can be generated by addressing them with tight-focused laser beams \cite{Weitenberg:2011}.
Additionally, we need to mention that the initial state with atoms on the $s$-band of the gauge sites can be lifted up to the corresponding $p$-band with the help of a band-mapping technique \cite{Wirth:2011}.

\textbf{\textit{Numerical benchmarks.---}}We now present simulations for adiabatic sweep and global quench dynamics, calculated with iMPS techniques \cite{Uli_review,Paeckel_review,mptoolkit}, using an algorithm based on the time-dependent variational principle \cite{Haegeman2011,Haegeman2013,Haegeman2016,Vanderstraeten2019,vumps}. We adopt the standard procedure of realizing $2$D models by employing a cylindrical geometry of an infinite axis (i.e., thermodynamic limit in the $x$-direction: $L_x\to\infty$) and with a finite circumference $L_y$ (number of matter sites in the QLM, equivalent to $2L_y$ sites in the BHM) in the $y$-direction, along which periodic boundary conditions are employed. Numerical details are provided in the SM~\cite{SM}.

\begin{figure}[t!]
	\centering
	\includegraphics[width=\linewidth]{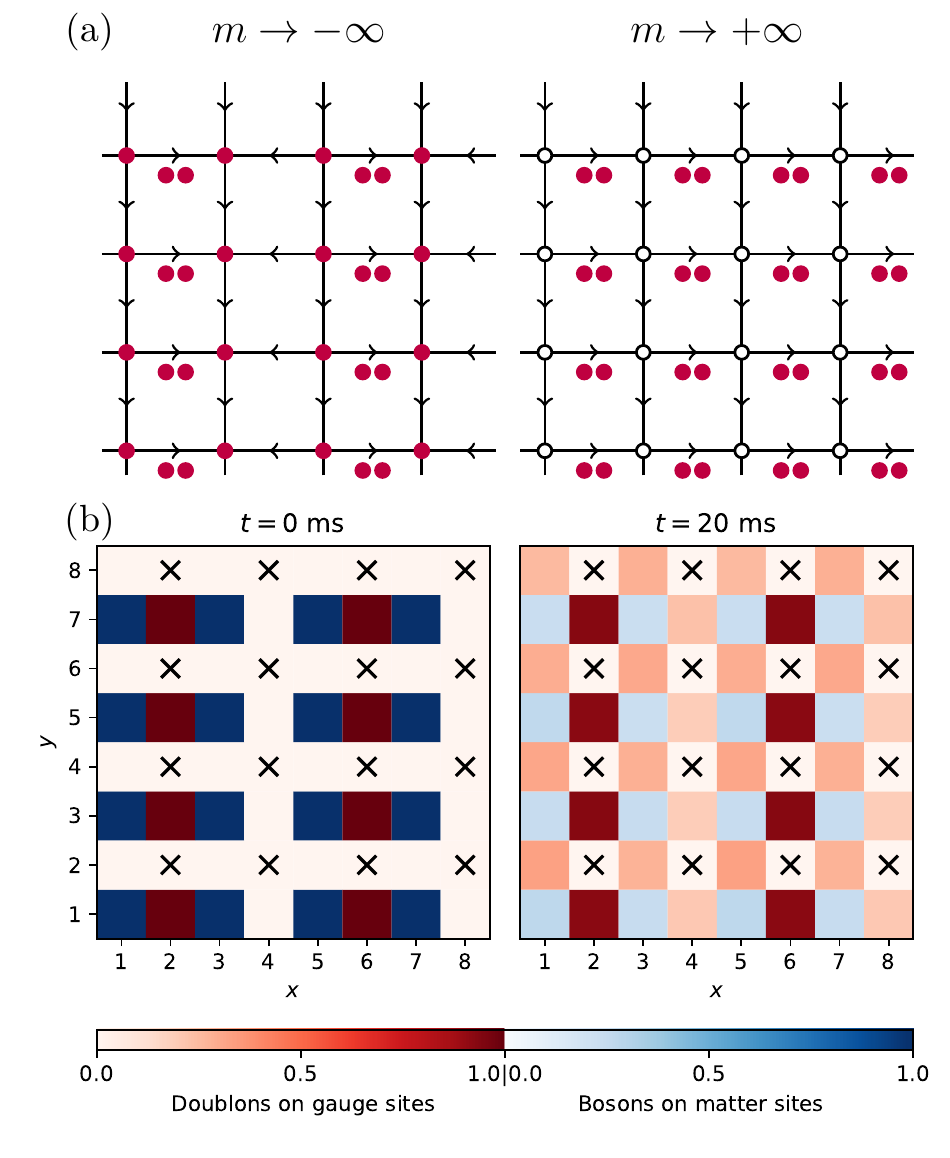}\\
	\includegraphics[width=\linewidth]{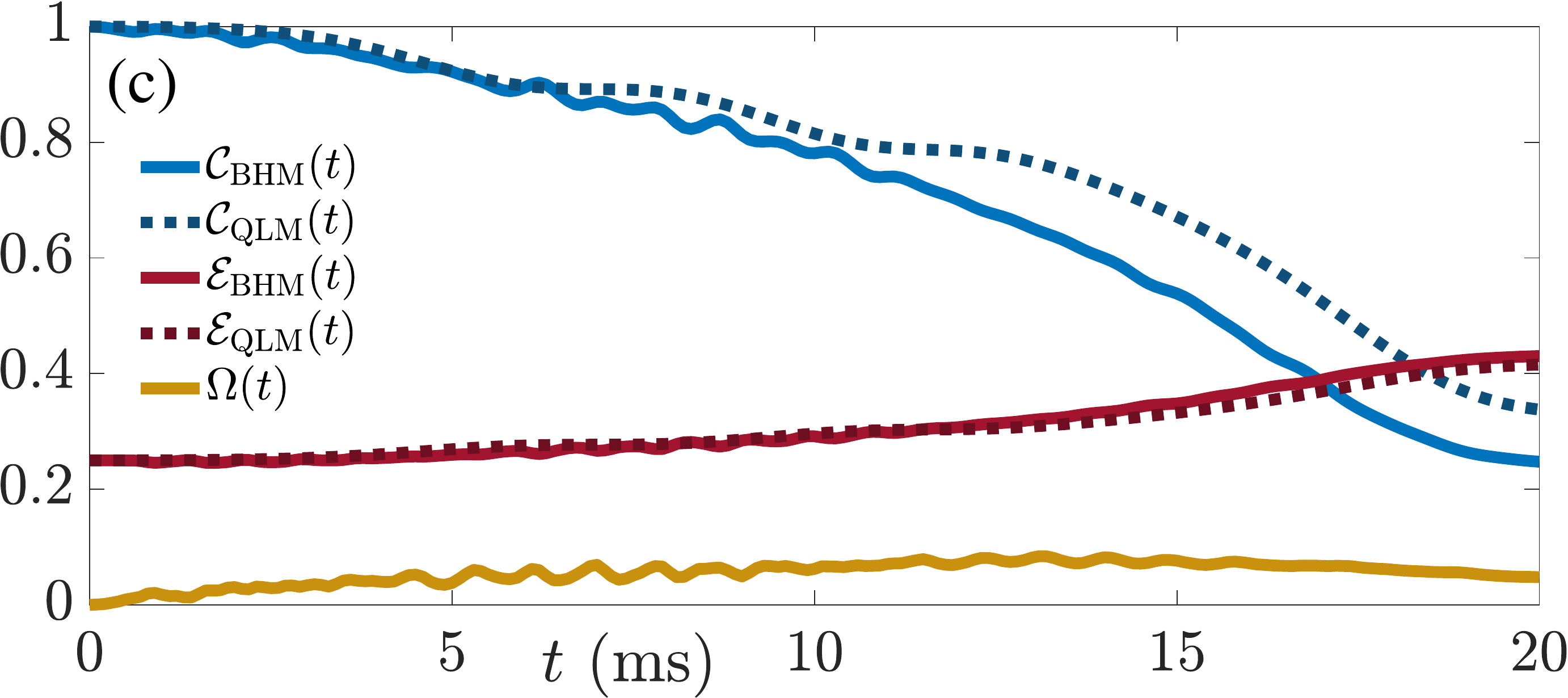}
	\caption{(Color online). Adiabatic sweep dynamics. (a) An exemplary charge-proliferated (left) and vacuum (right) state in the QLM and BHM representations. (b) Snapshots of the adiabatic sweep in the BHM~\eqref{eq:BHM} at $t=0$ ms, corresponding to the CP state in (a), and at $t=20$ ms, corresponding to the wave function at the end of the sweep. Crosses denote the forbidden sites on the BHM Lieb superlattice, and these sites take on even $j_x$ and even $j_y$ in our notation, while matter sites have odd $j_x$ and odd $j_y$, and gauge sites have one of the components $j_x$ and $j_y$ even and the second odd. (c) The corresponding dynamics of the chiral condensate~\eqref{eq:CC}, staggered electric flux~\eqref{eq:flux}, and the gauge violation~\eqref{eq:viol} in both the QLM and BHM, showing good agreement and a stabilized gauge invariance.}
	\label{fig:ramp}
\end{figure}

We first consider an adiabatic sweep where we start at $m/\kappa\to-\infty$ in a charge-proliferated (CP) state---depicted on the left in Fig.~\ref{fig:ramp}(a) within the QLM and BHM representations---in which every matter site is occupied, and slowly ramp the Hamiltonian parameters to $m/\kappa\to\infty$ in order to approach a vacuum state, where all matter sites are empty, an example of which is depicted on the right in Fig.~\ref{fig:ramp}(a). We set $L_y=4$ matter sites and fix $\eta=5\delta$. The exact ramps for the relevant parameters are within experimental validity and provided in the SM \cite{SM}. Figure~\ref{fig:ramp}(b) shows singlon (doublon) density maps on the matter (link) sites in the BHM simulation at $t=0$ and $t=20$ ms at the end of the sweep. The time-evolved wave function at $t=20$ ms is not an exact vacuum due to the finite speed of the ramp. The exemplary vacuum shown in Fig.~\ref{fig:ramp}(a) is not unique due to the degeneracy of the vacuum states, and so it is not surprising that the quantum simulation will end in a state close to a superposition of multiple vacua. Indeed, the density snapshot at $t=20$ ms shows that the initially unoccupied links at $t=0$ ms are now roughly equally occupied at $t=20$ ms, whereas the links that were initially fully occupied remain as such. 
To better benchmark the mapping, we look at the corresponding dynamics of the chiral condensate, defined in the QLM and BHM bases as
\begin{subequations}\label{eq:CC}
\begin{align}
    \mathcal{C}_\mathrm{QLM}(t)&=\lim_{L_x\to\infty}\frac{1}{L_xL_y}\sum_\mathbf{r}\bra{\psi(t)}\hat{\phi}_\mathbf{r}^\dagger\hat{\phi}_\mathbf{r}\ket{\psi(t)},\\
    \mathcal{C}_\mathrm{BHM}(t)&=\lim_{L_x\to\infty}\frac{1}{L_xL_y}\sum_{\mathbf{j}\in\text{matter}}\bra{\psi(t)}\hat{n}_\mathbf{j}\ket{\psi(t)},
\end{align}
\end{subequations}
respectively, and the staggered electric flux given by
\begin{subequations}\label{eq:flux}
\begin{align}
    \mathcal{E}_\mathrm{QLM}(t)&=\lim_{L_x\to\infty}\frac{1}{2L_xL_y}\sum_{\mathbf{r},\nu}\bra{\psi(t)}\hat{s}^z_{\mathbf{r},\mathbf{e}_\nu}\ket{\psi(t)},\\
    \mathcal{E}_\mathrm{BHM}(t)&=\lim_{L_x\to\infty}\frac{1}{2L_xL_y}\sum_{\mathbf{j}\in\text{gauge}}\bra{\psi(t)}\hat{n}^\mathrm{d}_\mathbf{j}\ket{\psi(t)},
\end{align}
\end{subequations}
where $\ket{\psi(t)}{=}\hat{\mathcal{T}}e^{-i\int_0^tds\hat{H}(s)}\ket{\psi_0}$, $\hat{H}(t)$ is either the QLM or BHM Hamiltonian at time $t$, $\ket{\psi_0}$ is the initial state, $\hat{\mathcal{T}}$ is the time-ordering operator, and $\hat{n}^\mathrm{d}_\mathbf{j}=\hat{b}_\mathbf{j}^\dagger\hat{b}_\mathbf{j}^\dagger\hat{b}_\mathbf{j}\hat{b}_\mathbf{j}/2$ is the doublon number operator. As shown in Fig.~\ref{fig:ramp}(c), the sweep dynamics of the chiral condensate in both the QLM and the BHM starts at unity and decreases until reaching at the end of the sweep a minimal finite value, with both models showing good agreement especially at early times. The staggered electric flux shows very good agreement between the QLM and BHM, starting off at $0.25$ at $t=0$, and approaching $\approx0.42$ at $t=20$ ms, which is smaller than its value of $0.5$ in the vacuum state. Quantitative differences between both models can be attributed to the small, albeit nonvanishing, gauge violation (orange curve), defined as the root mean square of the Gauss's-law operator,
\begin{align}\label{eq:viol}
    \Omega(t) = \lim_{L_x\rightarrow\infty} \sqrt{\frac{1}{L_xL_y} \sum_{\mathbf{r}} \bra{\psi(t)} \hat{G}_\mathbf{r} \ket{\psi(t)}^2}.
\end{align}
Throughout the whole dynamics we find that the gauge violation is restricted to below $8.4\%$. Importantly, we find that the gauge violation is not monotonically increasing, and shows a decrease towards the end of the sweep, highlighting the efficacy of the emergent gauge protection term in our implementation \cite{SM}.

\begin{figure}[t!]
	\centering
	\includegraphics[width=\linewidth]{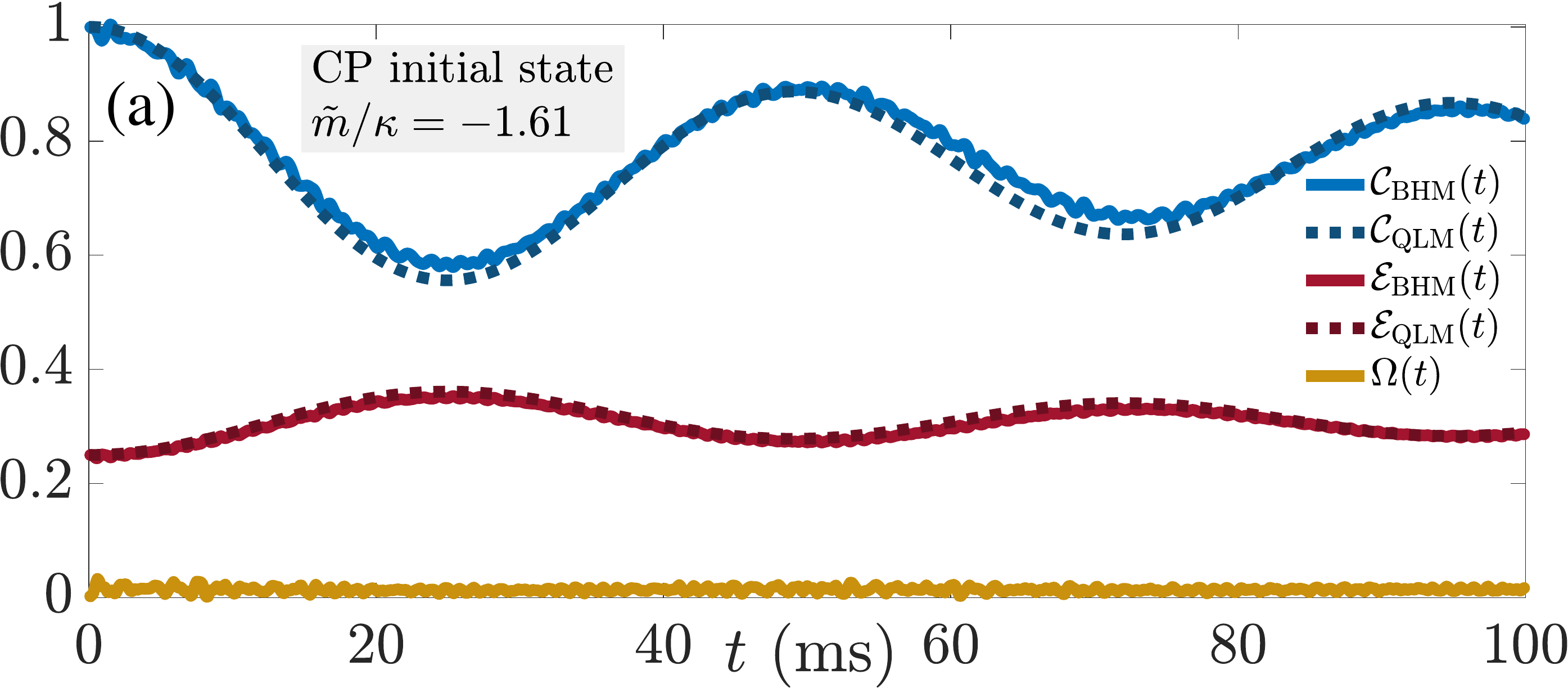}\\
	\vspace{1.1mm}
	\includegraphics[width=\linewidth]{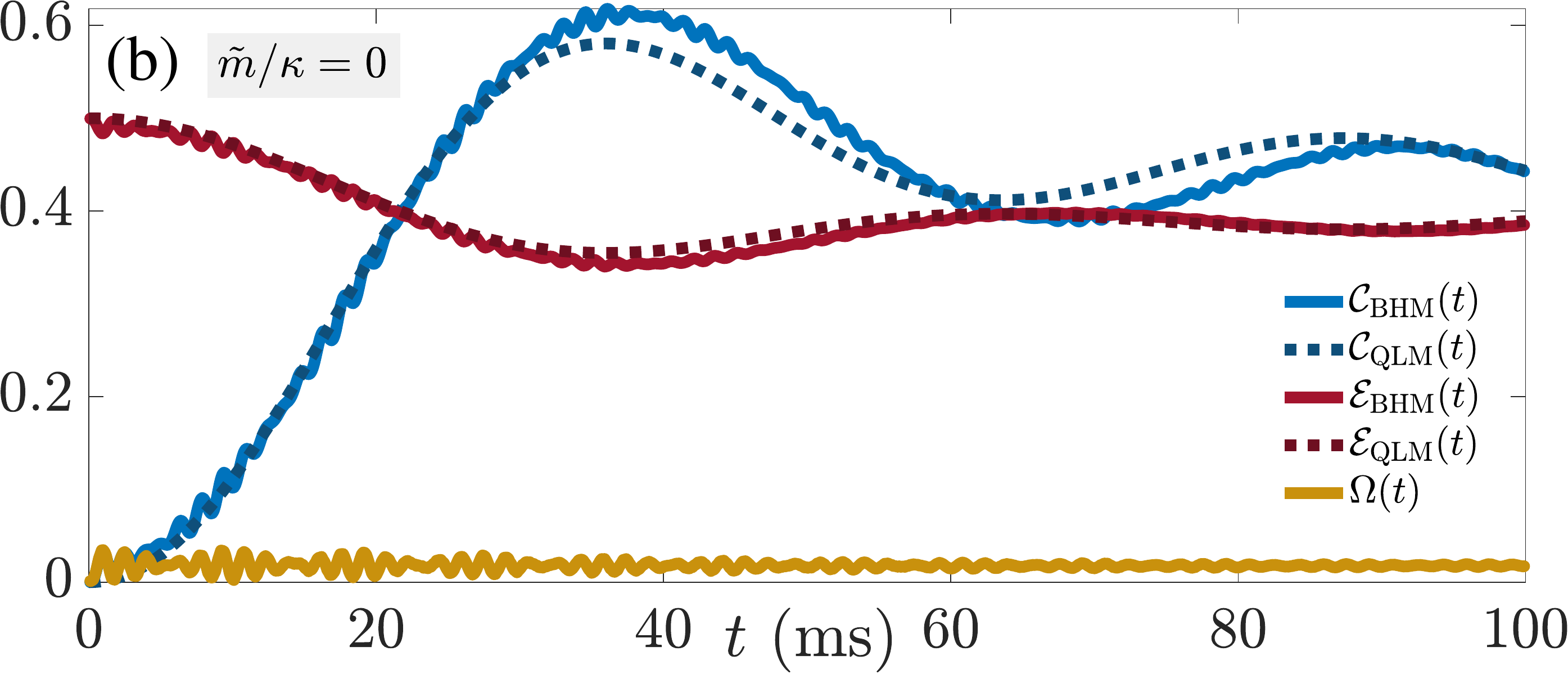}
	\caption{(Color online). Quench dynamics of the chiral condensate~\eqref{eq:CC}, staggered electric flux~\eqref{eq:flux}, and gauge violation~\eqref{eq:viol} in the wake of a global quench: (a) Starting in the CP state of Fig.~\ref{fig:ramp}(a) and quenching to $\tilde{m}/\kappa=-1.61$; (b) starting the vacuum initial state of Fig.~\ref{fig:ramp}(a) and quenching to $\tilde{m}/\kappa=0$. The results show impressive agreement between the QLM and BHM results, with gauge violation well-suppressed over all evolution times.}
	\label{fig:quench}
\end{figure}

Next, we calculate the dynamics of these observables in the wake of global quenches starting in the CP and vacuum states, shown in Fig.~\ref{fig:ramp}(a), with final mass values of $\tilde{m}/\kappa=-1.61$ and $0$, respectively. For these quenches, we fix $J=30$ Hz and $U=1300$ Hz, and $\delta$ is then determined based on the value of $\tilde{m}$ while again setting $\eta=5\delta$. Quench dynamics are numerically more costly than the adiabatic sweep. As such, due to computational overhead, we set $L_y=2$ matter sites ($4$ sites along the circumference in the case of the BHM). The corresponding dynamics is shown in Fig.~\ref{fig:quench}(a,b) for the CP and vacuum initial states, respectively. In both cases, we see very good agreement in the dynamics between the BHM and QLM up to all investigated evolution times. The gauge violation is impressively well-suppressed, always below $3.5\%$ in both cases.

\textbf{\textit{Outlook.---}}We have shown through second-order perturbation theory and numerical benchmarks using iMPS that the $2+1$D $\mathrm{U}(1)$ QLM can be faithfully mapped onto a two-dimensional Bose--Hubbard optical Lieb superlattice and quantum-simulated using spinless bosons. Although nontrivial adaptions are necessary to preserve gauge invariance in the $2+1$D case, this quantum simulation can be readily realized using already existing technology used in the $1+1$D case \cite{Yang2020,Zhou2022}. This would in principle allow the realization of a $2+1$D $\mathrm{U}(1)$ QLM on a $70\times70$ BHM, equivalent to $35$ matter sites along each direction. Despite impressive progress in classical computations in higher dimensions, in particular using tensor-network methods \cite{Magnifico2021,Hashizume2022}, this may enable quantum simulations that in the near future outperform  classical methods such as the one used in this work in terms of both size and maximal evolution times reached.

Such a quantum simulator of $2+1$D gauge theories opens the door to various exciting studies. For example, quantum many-body scars \cite{Turner2018,Moudgalya2018} are well-established in $1+1$D $\mathrm{U}(1)$ QLMs with dynamical matter \cite{Bernien2017,Surace2020,Desaules2022weak,Desaules2022prominent} and quantum link ladders without dynamical matter \cite{Banerjee2021}. It would be interesting to explore the fate of scars in $2+1$D $\mathrm{U}(1)$ QLMs with both dynamical matter and gauge fields, and further probe the connection to the quantum field theory limit suggested in $1+1$D \cite{Desaules2022weak,Desaules2022prominent}.

Another topic of great interest in gauge theories is that of thermalization, which is still an open issue in heavy-ion and electron-proton collisions, for example \cite{Berges_review}. First experimental works on the thermalization dynamics of gauge theories on a quantum simulator have been performed in $1+1$D \cite{Zhou2022}, and extending this to $2+1$D would shed more light on how gauge theories thermalize, and also allow for observing possible connections with entanglement spectra as theorized in recent work \cite{Mueller2022}. Furthermore, exotic topological and spin liquid phases recently theoretically demonstrated in $2+1$D gauge theories \cite{Cardarelli2017,Hashizume2022} would be interesting to probe on our proposed gauge-theory quantum simulator, as well as confinement \cite{Zohar2012,Halimeh2022tuning,Cheng2022tunable}.

\bigskip

\begin{acknowledgments}
J.C.H.~is grateful to Guo-Xian Su for stimulating discussions. 
This project has received funding from the European Research Council (ERC) under the European Union’s Horizon 2020 research and innovation programme (grant agreement No 804305). 
I.P.M.~acknowledges support from the Australian Research Council (ARC) Discovery Project Grants No.~DP190101515 and DP200103760. 
B.Y.~acknowledges support from National Key R$\&$D Program of China (grant 2022YFA1405800) and NNSFC (grant 12274199). 
P.H.~acknowledges support by the Google Research Scholar Award ProGauge, Provincia Autonoma di Trento, and Q@TN — Quantum Science and Technology in Trento. 
J.C.H.~acknowledges funding from the European Research Council (ERC) under the European Union’s Horizon 2020 research and innovation programm (Grant Agreement no 948141) — ERC Starting Grant SimUcQuam, and by the Deutsche Forschungsgemeinschaft (DFG, German Research Foundation) under Germany's Excellence Strategy -- EXC-2111 -- 390814868.
Numerical simulations were performed on The University of Queensland's School of Mathematics and Physics Core Computing Facility ``getafix''.
\end{acknowledgments}

\clearpage
\pagebreak
\newpage
\setcounter{equation}{0}
\setcounter{figure}{0}
\setcounter{table}{0}
\setcounter{page}{1}
\makeatletter
\renewcommand{\bibnumfmt}[1]{[S#1]}
\renewcommand{\citenumfont}[1]{S#1}
\renewcommand{\theequation}{S\arabic{equation}}
\renewcommand{\thefigure}{S\arabic{figure}}
\renewcommand{\thetable}{S\Roman{table}}
\renewcommand{\bibnumfmt}[1]{[S#1]}
\renewcommand{\citenumfont}[1]{S#1}
\widetext
\begin{center}
\textbf{--- Supplemental Material ---\\Large-Scale $2{+}1$D $\mathrm{U}(1)$ Gauge Theory with Dynamical Matter in a Cold-Atom Quantum Simulator}
\text{Jesse Osborne, Ian P.~McCulloch, Bing Yang, Philipp Hauke, and Jad C.~Halimeh}
\end{center}
\date{\today}
\maketitle
\tableofcontents
\section{$2+1$D Abelian Higgs model}
The $2+1$D lattice Abelian Higgs model is described by the Hamiltonian
\begin{align}\label{eq:QLMph}
    \hat{H}_\mathrm{QLM}=&-\kappa\sum_{\mathbf{r},\nu=x,y}c_{\mathbf{r},\mathbf{e}_\nu}\Big(\hat{\phi}_\mathbf{r}^\dagger\hat{s}^+_{\mathbf{r},\mathbf{e}_\nu}\hat{\phi}_{\mathbf{r}+\mathbf{e}_\nu}+\text{H.c.}\Big)+m\sum_\mathbf{r}c_\mathbf{r}\hat{\phi}^\dagger_\mathbf{r}\hat{\phi}_\mathbf{r}-\Lambda\sum_\mathbf{r}\Big(\hat{U}_{\square_\mathbf{r}}+\hat{U}^\dagger_{\square_\mathbf{r}}\Big),
\end{align}
where $\mathbf{r}{=}\big(r_x,r_y\big)^\intercal$ is the vector specifying the position of a lattice site, $\mathbf{e}_\nu$ is a unit vector along the direction $\nu{\in}\{x,y\}$ (lattice spacing is set to unity), $\kappa$ is the coupling strength, $c_{\mathbf{r},\mathbf{e}_x}{=}1$, $c_{\mathbf{r},\mathbf{e}_y}{=}(-1)^{r_x}$, and $c_\mathbf{r}{=}(-1)^{r_x{+}r_y}$. The hard-core bosonic annihilation and creation operators $\hat{\phi}_\mathbf{r}$ and $\hat{\phi}_\mathbf{r}^\dagger$, respectively, act on the matter field at site $\mathbf{r}$ with mass $m$, while the spin-$1/2$ operators $\hat{s}^\pm_{\mathbf{r},\mathbf{e}_\nu}$ and $\hat{s}^z_{\mathbf{r},\mathbf{e}_\nu}$ represent the gauge and electric fields, respectively, at the link between sites $\mathbf{r}$ and $\mathbf{r}{+}\mathbf{e}_\nu$, and the plaquette term $\hat{U}_{\square_\mathbf{r}}{=}\hat{s}^+_{\mathbf{r},\mathbf{e}_x}\hat{s}^+_{\mathbf{r}+\mathbf{e}_x,\mathbf{e}_y}\hat{s}^-_{\mathbf{r}+\mathbf{e}_y,\mathbf{e}_x}\hat{s}^-_{\mathbf{r},\mathbf{e}_y}$ governs the magnetic interactions for gauge fields. The tunneling and mass terms are staggered as per the Kogut--Susskind formulation \cite{Kogut1975-S}. The generator of the $\mathrm{U}(1)$ gauge symmetry of Hamiltonian~\eqref{eq:QLM} is
\begin{align}\label{eq:Grph}
    \hat{G}_\mathbf{r}=\hat{\phi}^\dagger_\mathbf{r}\hat{\phi}_\mathbf{r}{-}\frac{1{-}(-1)^{r_x+r_y}}{2}{-}\sum_{\nu=x,y}\Big(\hat{s}^z_{\mathbf{r},\mathbf{e}_\nu}{-}\hat{s}^z_{\mathbf{r}-\mathbf{e}_\nu,\mathbf{e}_\nu}\Big),
\end{align}
which is equivalent to a discretized version of Gauss's law. Since the $\mathrm{U}(1)$ gauge symmetry is Abelian, the commutation relations $\big[\hat{G}_\mathbf{r},\hat{G}_{\mathbf{r}'}\big]=0$ are satisfied. The gauge invariance of Eq.~\eqref{eq:QLMph} is encoded in the commutation relations $\big[\hat{H}_\mathrm{QLM},\hat{G}_\mathbf{r}\big]=0,\,\forall \mathbf{r}$.

In this proof-of-principle study, we neglect the plaquette term that governs magnetic interactions for gauge fields, and so we set $\Lambda{=}0$. Employing the particle-hole transformations
\begin{subequations}
\begin{align}
    &\hat{\phi}_\mathbf{r}\to\frac{1+(-1)^{r_x+r_y}}{2}\hat{\phi}_\mathbf{r}+\frac{1-(-1)^{r_x+r_y}}{2}\hat{\phi}_\mathbf{r}^\dagger,\\
    &\hat{s}^+_{\mathbf{r},\mathbf{e}_\nu}\to\frac{(-1)^{r_x}+(-1)^{r_y}}{2}\hat{s}^-_{\mathbf{r},\mathbf{e}_\nu}+\frac{(-1)^{r_x}-(-1)^{r_y}}{2}\hat{s}^+_{\mathbf{r},\mathbf{e}_\nu},
\end{align}
\end{subequations}
we arrive at Eqs.~\eqref{eq:QLM} and~\eqref{eq:Gr} in the main text, which not only rid us of the staggering coefficients of Eq.~\eqref{eq:QLMph}, but also connect more intuitively to the bosonic mapping outlined in this work (see Fig.~\ref{fig:setup}).

\section{Mapping onto two-dimensional Bose--Hubbard superlattice}

The $2+1$D $\mathrm{U}(1)$ QLM~\eqref{eq:QLM} can be mapped onto the two-dimensional Bose--Hubbard superlattice with Hamiltonian
~\eqref{eq:BHM} given in the main text, and which can in turn be realized in a cold-atom quantum simulator with existing technology \cite{Yang2020-S,Zhou2022-S} (see discussion in the main text).

We will now explain this mapping in detail, which is based on earlier work in $1+1$D \cite{Yang2020-S}. On matter sites, we restrict the local Hilbert space to $\mathcal{H}_\mathbf{r}{=}\mathrm{span}\{\ket{0}_\mathbf{r},\ket{1}_\mathbf{r}\}$, i.e., the allowed matter configurations are represented by no bosons (empty) or a single boson (occupied). On gauge sites, we require the local Hilbert space to be $\mathcal{H}_{\mathbf{r},\mathbf{e}_\nu}{=}\mathrm{span}\{\ket{0}_{\mathbf{r},\mathbf{e}_\nu},\ket{2}_{\mathbf{r},\mathbf{e}_\nu}\}$. If the gauge site represents a horizontal (vertical) link in the corresponding QLM, then zero occupation represents a leftward (downward) electric flux polarization, while a doublon occupation represents a rightward (upward) electric flux polarization; see Fig.~\ref{fig:setup}(b). This can be formulated in the mappings

\begin{subequations}\label{eq:mapping}
\begin{align}
    &\hat{\phi}_\mathbf{r}=\hat{\mathcal{P}}_\mathbf{r}\hat{b}_\mathbf{r}\hat{\mathcal{P}}_\mathbf{r},\,\,\,\,\,\,\,
    \hat{\phi}^\dagger_\mathbf{r}=\hat{\mathcal{P}}_\mathbf{r}\hat{b}^\dagger_\mathbf{r}\hat{\mathcal{P}}_\mathbf{r},\,\,\,\,\,\,\,
    \hat{\phi}^\dagger_\mathbf{r}\hat{\phi}_\mathbf{r}=\hat{\mathcal{P}}_\mathbf{r}\hat{b}^\dagger_\mathbf{r}\hat{b}_\mathbf{r}\hat{\mathcal{P}}_\mathbf{r},\\
    &\hat{s}^-_{\mathbf{r},\mathbf{e}_\nu}=\frac{1}{\sqrt{2}}\hat{\mathcal{P}}_{\mathbf{r},\mathbf{e}_\nu}\big(\hat{b}_{\mathbf{r},\mathbf{e}_\nu}\big)^2\hat{\mathcal{P}}_{\mathbf{r},\mathbf{e}_\nu},\,\,\,\,\,\,\,
    \hat{s}^+_{\mathbf{r},\mathbf{e}_\nu}=\frac{1}{\sqrt{2}}\hat{\mathcal{P}}_{\mathbf{r},\mathbf{e}_\nu}\big(\hat{b}^\dagger_{\mathbf{r},\mathbf{e}_\nu}\big)^2\hat{\mathcal{P}}_{\mathbf{r},\mathbf{e}_\nu},\,\,\,\,\,\,\,
    \hat{s}^z_{\mathbf{r},\mathbf{e}_\nu}=\frac{1}{2}\hat{\mathcal{P}}_{\mathbf{r},\mathbf{e}_\nu}\big(\hat{b}_{\mathbf{r},\mathbf{e}_\nu}^\dagger\hat{b}_{\mathbf{r},\mathbf{e}_\nu}-1\big)\hat{\mathcal{P}}_{\mathbf{r},\mathbf{e}_\nu},
\end{align}
\end{subequations}
where $\hat{\mathcal{P}}_\mathbf{r}$ and $\hat{\mathcal{P}}_{\mathbf{r},\mathbf{e}_\nu}$ are local projectors onto the local Hilbert spaces $\mathcal{H}_\mathbf{r}$ and $\mathcal{H}_{\mathbf{r},\mathbf{e}_\nu}$ of the matter and gauge sites, respectively. Note that, for now, we have employed the QLM indexing for the bosonic operators of the BHM. Plugging Eqs.~\eqref{eq:mapping} into Eq.~\eqref{eq:QLM}, we obtain
\begin{align}\label{eq:Hbos}
    \hat{H}_\text{b}{=}\hat{\mathcal{P}}\sum_\mathbf{r}\bigg[\frac{\kappa}{\sqrt{2}}\sum_{\nu=x,y}\big(\hat{b}_\mathbf{r}^\dagger\hat{b}_{\mathbf{r},\mathbf{e}_\nu}^2\hat{b}_{\mathbf{r}+\mathbf{e}_\nu}^\dagger{+}\text{H.c.}\big){+}m\hat{b}_\mathbf{r}^\dagger\hat{b}_\mathbf{r}\bigg]\hat{\mathcal{P}},
\end{align}
where $\hat{\mathcal{P}}=\prod_{\mathbf{r},\nu}\hat{\mathcal{P}}_\mathbf{r}\hat{\mathcal{P}}_{\mathbf{r},\mathbf{e}_\nu}$. Hamiltonian $\hat{H}_\text{b}$ can be mapped onto an effective model derived from the BHM~\eqref{eq:BHM} through degenerate perturbation theory~\cite{Yang2020-S} in the regime of $\eta\gg U\sim2\delta\gg J,m$, with $\kappa\approx4\sqrt{2}J^2/U$. In this limit, the hopping term $\propto J$ in the BHM becomes a perturbation to the diagonal terms described by the Hamiltonian
\begin{align}\label{eq:Hdiag}
    \hat{H}_\text{diag}=\sum_\mathbf{r}\bigg\{\alpha\frac{U}{2}\hat{n}_\mathbf{r}\big(\hat{n}_\mathbf{r}-1\big)+\vec{\gamma}^\intercal\mathbf{p}_\mathbf{r}\hat{n}_\mathbf{r}+\sum_{\nu=x,y}\bigg[\frac{U}{2}\hat{n}_{\mathbf{r},\mathbf{e}_\nu}\big(\hat{n}_{\mathbf{r},\mathbf{e}_\nu}-1\big)-\delta\hat{n}_{\mathbf{r},\mathbf{e}_\nu}+\vec{\gamma}^\intercal\big(\mathbf{p}_\mathbf{r}+\mathbf{e}_\nu\big)\hat{n}_{\mathbf{r},\mathbf{e}_\nu}\bigg]\bigg\},
\end{align}
where $\mathbf{p}_\mathbf{r}=(2r_x-1,2r_y-1)^\intercal$ maps the location of matter sites, and hence gauge links, from the spatial coordinates of the QLM lattice to those of the BHM lattice.\footnote{Recall from Fig.~\ref{fig:ramp}(b) that on the BHM superlattice, a matter site of index $\mathbf{j}$ has both $j_x$ and $j_y$ odd, while a gauge site of index $\mathbf{j}$ has one of $j_x$ and $j_y$ odd and the other even, while a forbidden site with index $\mathbf{j}$ has both $j_x$ and $j_y$ even.} One can then derive a ``proto'' Gauss's law with generator
\begin{align}\label{eq:G_proto}
    \hat{\mathcal{G}}_\mathbf{r}=(-1)^\mathbf{r}\bigg[\hat{n}_\mathbf{r}+\frac{1}{2}\sum_{\nu=x,y}\hat{n}_{\mathbf{r},\pm\mathbf{e}_\nu}\big(\hat{n}_{\mathbf{r},\pm\mathbf{e}_\nu}{-}1\big)-2\bigg],
\end{align}
which, along with $\delta=m+U/2$ and an inconsequential energy constant, can be plugged into Eq.~\eqref{eq:Hdiag} as
\begin{align}\label{eq:Hdiag_proto}
    \hat{H}_\text{diag}=\sum_\mathbf{r}\bigg\{\alpha\frac{U}{2}\hat{n}_\mathbf{r}\big(\hat{n}_\mathbf{r}-1\big)+\sum_{\nu=x,y}\bigg[\frac{U}{2}\hat{n}_{\mathbf{r},\mathbf{e}_\nu}\big(\hat{n}_{\mathbf{r},\mathbf{e}_\nu}-2\big)-m\hat{n}_{\mathbf{r},\mathbf{e}_\nu}+\vec{\gamma}^\intercal\mathbf{e}_\nu\hat{n}_{\mathbf{r},\mathbf{e}_\nu}\bigg]+\omega_\mathbf{r}\hat{\mathcal{G}}_\mathbf{r}\bigg\},
\end{align}
where $\omega_\mathbf{r}=(-1)^\mathbf{r}\vec{\gamma}^\intercal\mathbf{p}_\mathbf{r}$. As such, we see that the gauge-invariant diagonal Hamiltonian $\hat{H}_\text{diag}$ includes a \textit{Stark gauge protection} term $\sum_\mathbf{r}\omega_\mathbf{r}\hat{\mathcal{G}}_\mathbf{r}$ that stabilizes the gauge theory against the gauge-noninvariant processes due to the tunneling term in the BHM \cite{Halimeh2020e-S,Lang2022stark-S}.

\section{Undesired processes in second-order perturbation theory}\label{sec:PT}
\begin{figure}[t!]
	\centering
	\includegraphics[width=0.48\linewidth]{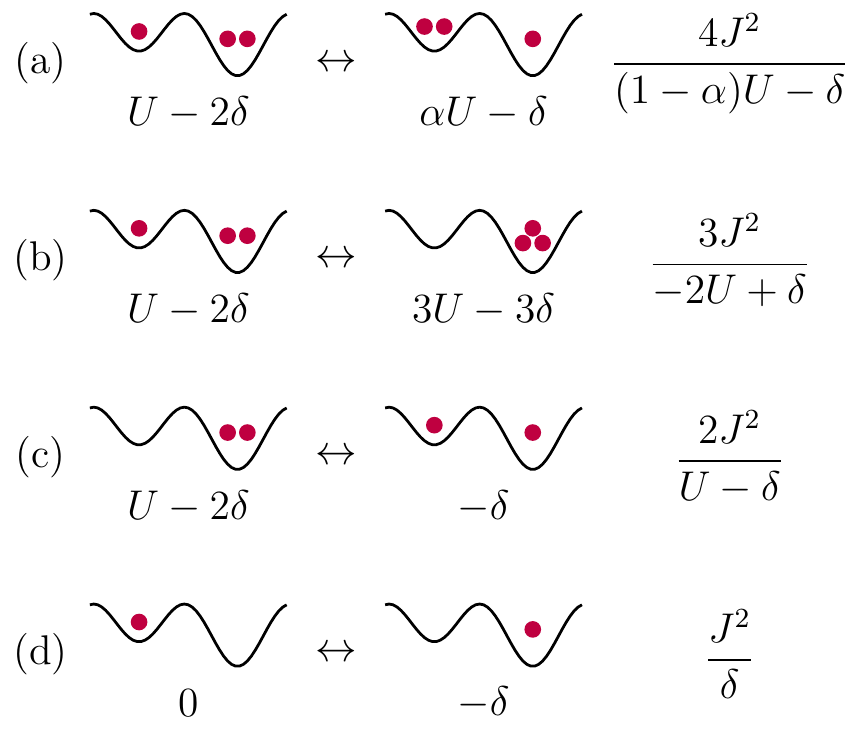}
	\caption{(Color online). Undesired gauge-invariant processes in second-order perturbation theory in the mapping of the $2+1$D $\mathrm{U}(1)$ QLM onto a two-dimensional BHM optical superlattice. Processes (a) and (b) are much less probable in $1+1$D since in the gauge-invariant target sector an unoccupied matter site is adjacent to an empty gauge site and one with a doublon, while an occupied matter site is adjacent to two empty gauge sites, corresponding to the processes in (c) and (d), respectively. All these processes lead to a slightly renormalization $m\to\tilde{m}$ according to Eq.~\eqref{eq:mtilde} in the $2+1$D $\mathrm{U}(1)$ QLM; see also Fig.~\ref{fig:ramp_protocol}(d).}
	\label{fig:PerturbationTheory}
\end{figure}
In second-order perturbation theory in the hopping term, there are some processes in the Bose--Hubbard model which do not correspond to any process in the QLM, but will effectively renormalize the mass term in the QLM in the zero-gauge sector.
These processes correspond to the matter-gauge site configurations shown in Fig.~\ref{fig:PerturbationTheory}.\footnote{There are also other processes where bosons on matter sites may tunnel across gauge sites, but these are off-resonant due to the tilt $\vec{\gamma}$.}
Now since we restrict ourselves to the gauge sector with zero background charge, according to Gauss's law~\eqref{eq:G_proto}, each occupied matter site has exactly one doublon on the surrounding gauge sites, and three unoccupied gauge sites, while each unoccupied gauge site has exactly two doublons on the surrounding gauge sites.
Therefore, the transformed QLM~\eqref{eq:QLM} is perturbed by (neglecting the minimal effect of the tilt $\vec{\gamma}$)
\begin{align}
    \hat{H}_\text{pert} = \sum_\mathbf{r} \Bigg\{ \left[ \frac{4J^2}{(1-\alpha)U-\delta} + \frac{3J^2}{\delta-2U} \right] \hat{\phi}^\dag_\mathbf{r} \hat{\phi}_\mathbf{r}+ \frac{4J^2}{U-\delta} \left( 1 - \hat{\phi}^\dag_\mathbf{r} \hat{\phi}_\mathbf{r} \right)
    + \frac{3J^2}{\delta} \hat{\phi}^\dag_\mathbf{r} \hat{\phi}_\mathbf{r}\Bigg\}.
\end{align}
Apart from shifting the total energy by a constant, this will result in the same Hamiltonian~\eqref{eq:QLM} with the renormalized mass
\begin{align}\label{eq:mtilde}
    \tilde{m} = m + J^2 \left[ \frac{4}{(1-\alpha)U-\delta} - \frac{3}{2U-\delta} - \frac{4}{U-\delta} + \frac{3}{\delta} \right].
\end{align}

We note that although these processes have a nontrivial effect on the mapping of the $2+1$D QLM, in the $1+1$D case~\cite{Yang2020-S}, the effect is negligible, as we shall now show.
According to the mapping of Gauss's law in $1+1$D, each occupied matter site has two unoccupied gauge sites on either side of it [corresponding to the process in Fig.~\ref{fig:PerturbationTheory}(d)], and each unoccupied matter site has only one doublon on the neighboring gauge sites [the process in Fig.~\ref{fig:PerturbationTheory}(c)].
Therefore, the perturbation Hamiltonian is
\begin{align}
    \hat{H}_\text{pert} = \sum_j \left[
    \frac{2J^2}{\delta} \hat{\phi}^\dag_j \hat{\phi}_j
    + \frac{2J^2}{U-\delta} \left( 1 - \hat{\phi}^\dag_j \hat{\phi}_j \right)
    \right],
\end{align}
and so the mass is renormalized by
\begin{align}
    \tilde{m} = m + 2J^2 \left[ \frac{1}{\delta} - \frac{1}{U-\delta} \right].
\end{align}
Since we typically work in the region where $U \approx 2\delta$, the renormalization term will be negligibly small.

\section{Numerical details}\label{sec:iMPS}
We use infinite matrix product state (iMPS) numerical methods~\cite{paeckel2019-S,mptoolkit-S} to simulate the time evolution for both the Bose--Hubbard and quantum link models.
We use the standard procedure for using iMPSs to represent 2D systems by using a cylindrical geometry, where the finite circumference of the cylinder is $L_y$ matter sites ($2L_y$ sites in the BHM), and the system is infinitely long in the $x$ direction, but the state is invariant by translations of two matter sites.
All of the simulations use either the CP state ($m \rightarrow -\infty$) or the vaccum state ($m \rightarrow \infty$), using the configurations shown in Fig.~\ref{fig:ramp}(a), which can be written as product states in terms of the local bases, and so no special procedure is needed to generate the initial states.

To perform the time evolution simulations, we use a version of the time-dependent variational principle (TDVP)~\cite{Haegeman2011-S,Haegeman2016-S}, adapted for infinite systems with a large MPS unit cell.
This is done by first solving the fixed points for the Hamiltonian environments~\cite{michel2010-S}, and then sweeping across the unit cell as in standard finite TDVP~\cite{Haegeman2016-S}.
After the first left (right) sweep we update the right (left) Hamiltonian environment and sweep the original unit cell again until the unit cell fidelity between sweeps is sufficiently close to one.
In practice, for the simulations considered in this paper and for a reasonably small timestep, we usually reach the desired threshold after the second sweep, so there is no need to perform a large number of sweeps over the same unit cell to perform a single timestep.
We use a single-site evolution scheme with adaptive bond dimension expansion~\cite[App.~B]{vumps-S}.
We use a timestep of 0.1\,ms, and a maximum on-site boson occupation of 3.

Because of the difficulty in simulating a Hamiltonian with a linear tilt on a translation-invariant cylinder with a finite circumference, we transform to the ``dynamic gauge''~\cite{zisling2022-S}, where the linear tilts in the Hamiltonian in the $x$ and $y$ directions are transformed into a time-dependent complex phase in the hopping terms in the same direction.
In the dynamic gauge, the Bose--Hubbard Hamiltonian \eqref{eq:BHM} takes the form
\begin{align}\label{eq:BHM-dynamic-gauge}
\hat{H}_\text{BHM}(t)=\sum_\mathbf{j}\bigg[&J\sum_{\nu=x,y}\Big(\mathrm{e}^{\mathrm{i}\gamma_\nu t} \hat{b}_\mathbf{j}^\dagger\hat{b}_\mathbf{j+\mathbf{e}_\nu}+\text{H.c.}\Big)+\frac{U_\mathbf{j}}{2}\hat{n}_\mathbf{j}\big(\hat{n}_\mathbf{j}-1\big)-\big(\delta_\mathbf{j}+\eta_\mathbf{j}\big)\hat{n}_\mathbf{j}\bigg].
\end{align}

\begin{figure}[t!]
	\centering
	\includegraphics[width=0.34\linewidth]{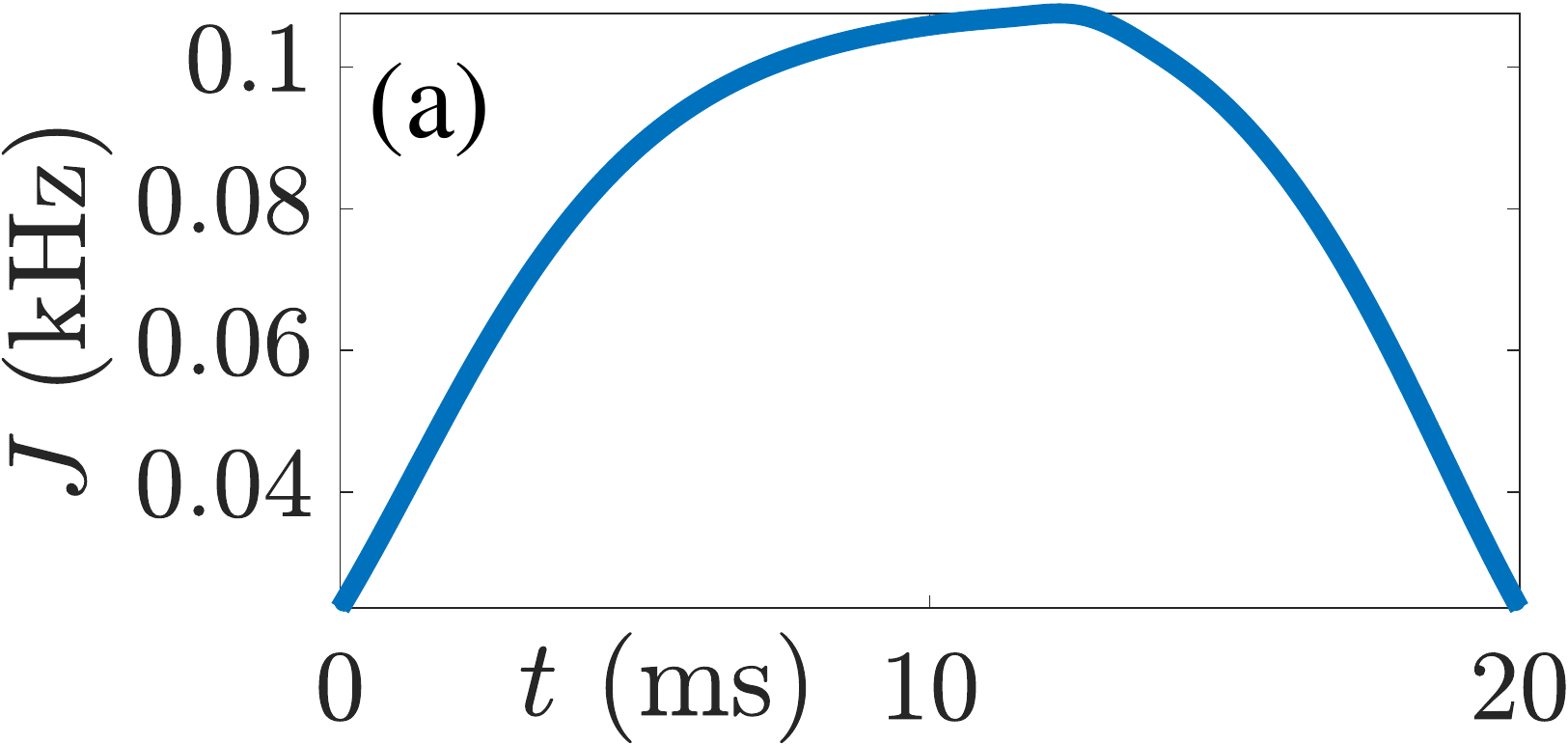}\quad\includegraphics[width=0.34\linewidth]{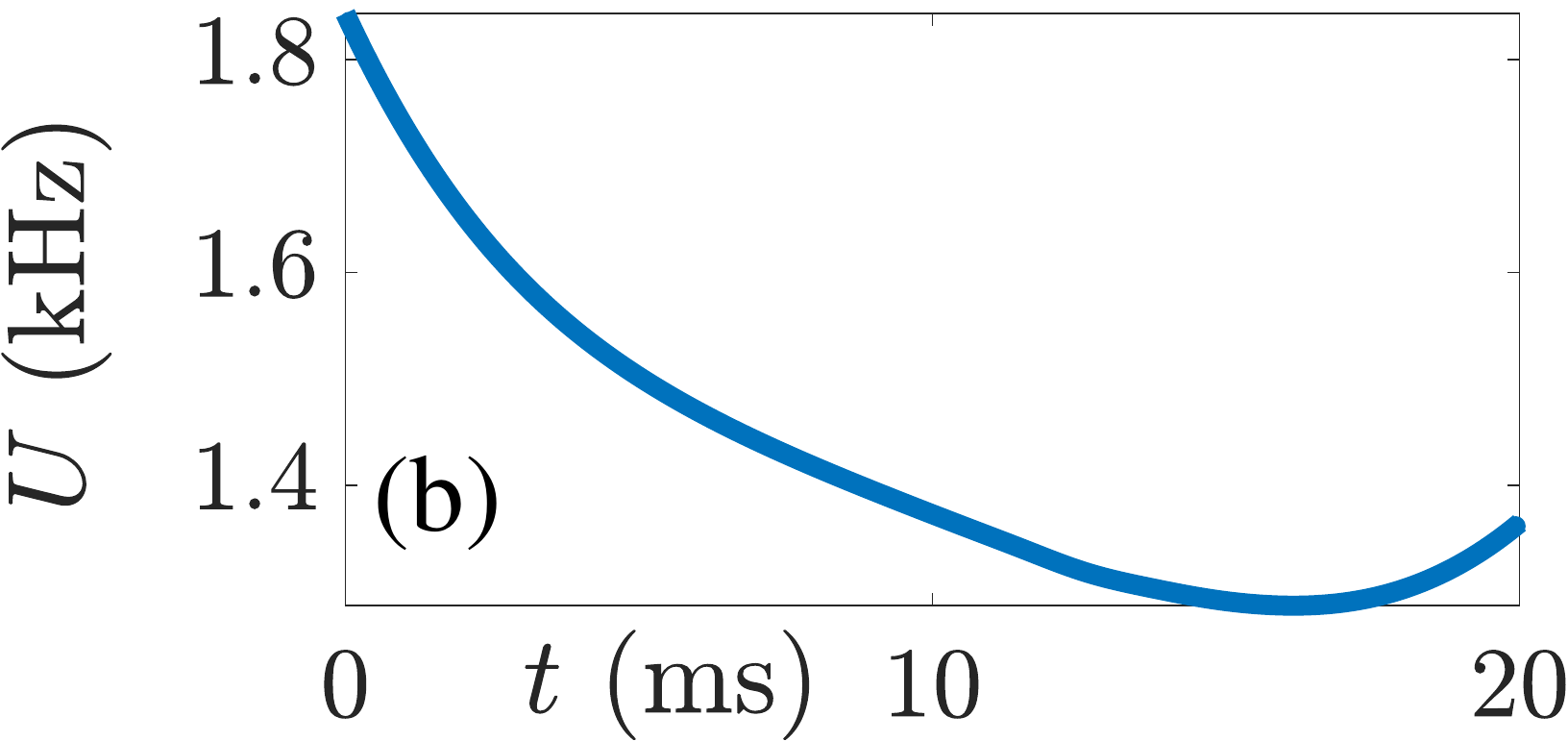}\\
	\vspace{1.1mm}
	\includegraphics[width=0.34\linewidth]{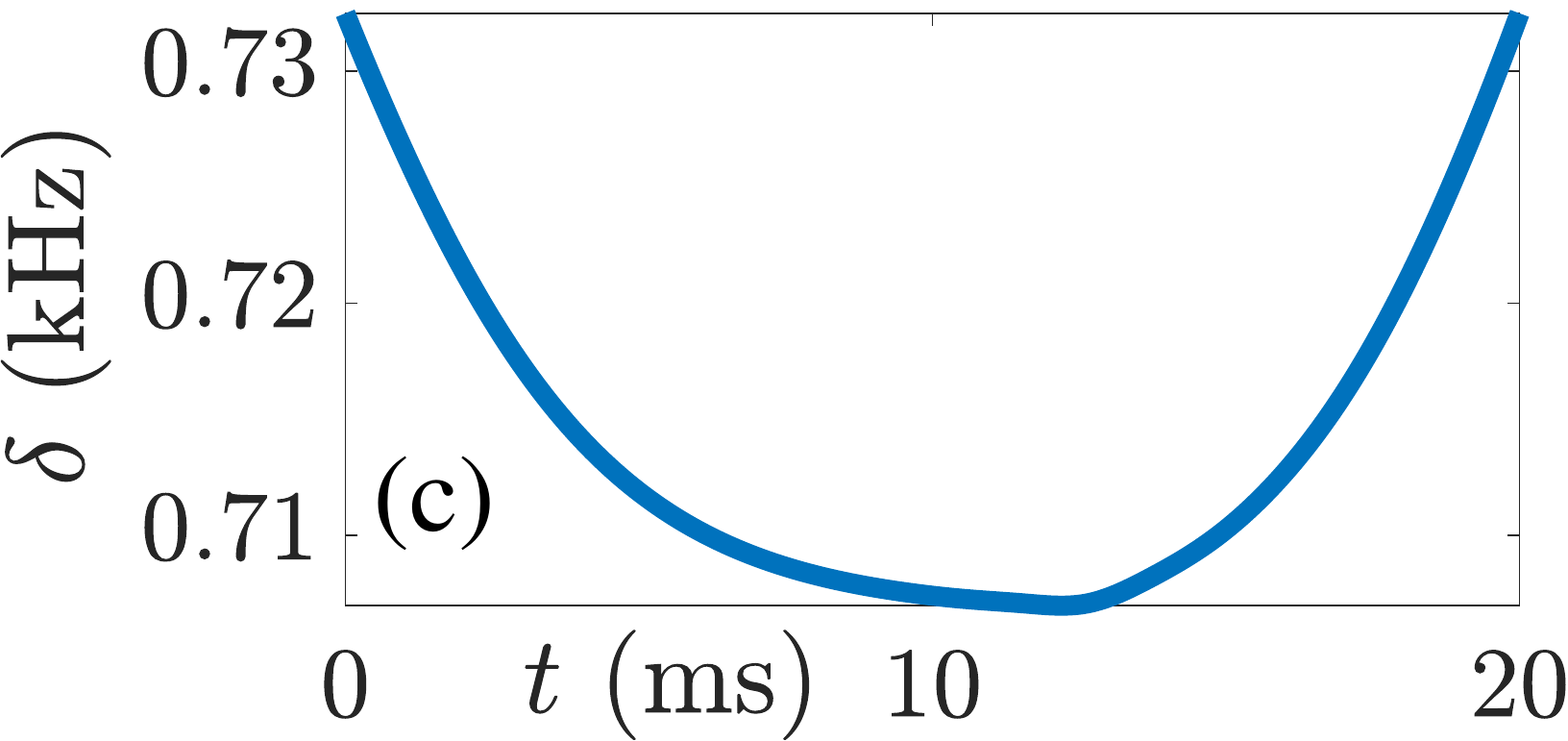}\quad\includegraphics[width=0.34\linewidth]{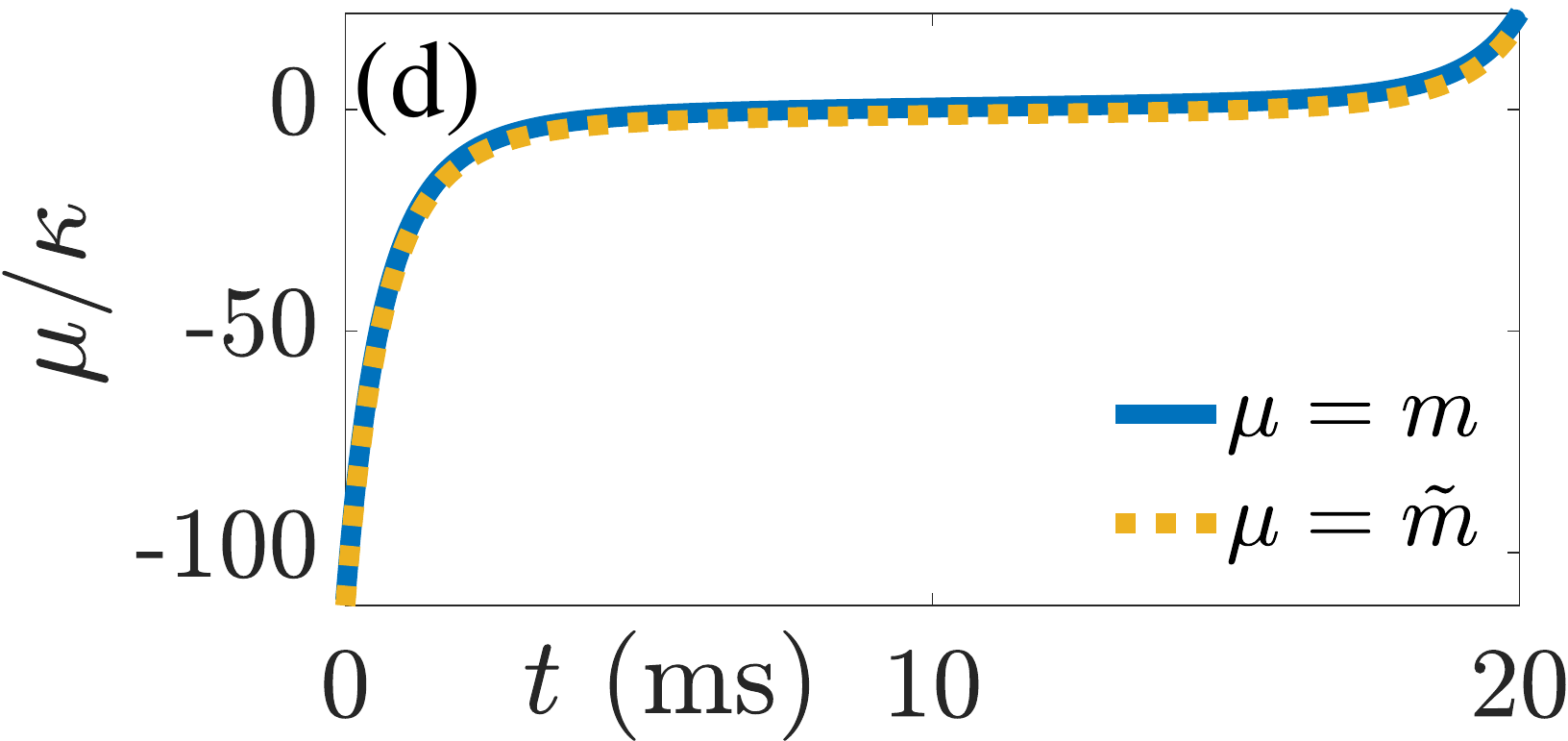}
	\caption{(Color online). Ramp protocol for the parameters of the BHM~\eqref{eq:BHM} used for the adiabatic sweep of Fig.~\ref{fig:ramp}: (a) The tunneling strength $J$, (b) the on-site interaction $U$, (c) the staggering potential $\delta$, (d) and the resulting (renormalized) mass, where the mass is $m=\delta-U/2$, the renormalized mass $\tilde{m}$ is given by Eq.~\eqref{eq:mtilde}, and $\kappa\approx4\sqrt{2}J^2/U$ as obtained from degenerate perturbation theory.}
	\label{fig:ramp_protocol}
\end{figure}

\section{Ramp protocol for BHM parameters}
For the adiabatic sweep shown in Fig.~\ref{fig:ramp} of the main text, we have adopted a ramp protocol that has been used in a cold-atom quantum simulator of the $1+1$D $\mathrm{U}(1)$ QLM \cite{Yang2020-S,Zhou2022-S}. This ramp protocol in the BHM parameters $J$, $U$, and $\delta$ is shown in Fig.~\ref{fig:ramp_protocol}. Of course, other ramp protocols are possible, but we have opted to use this one to highlight the experimental feasibility of our work in that already existing setups can be extended to probe our findings. It is worth noting how the renormalized mass $\tilde{m}$ is qualitatively the same as $m$ during the ramp protocol, and only slightly differs from it quantitatively; see Fig.~\ref{fig:ramp_protocol}(d).

\end{document}